\documentclass[letterpaper,preprintnumbers,prd,twocolumn,nofootinbib,nobibnotes,showpacs]{revtex4}
\usepackage{amsfonts}
\usepackage{mathrsfs}
\usepackage{epsfig}
\usepackage{graphicx}%
\usepackage{dcolumn}
\usepackage{amsmath}

\makeatletter
\def\btt#1{\texttt{\@backslashchar#1}}%
\DeclareRobustCommand\bblash{\btt{\@backslashchar}}%
\makeatother
\begin{document}

\title{Ghosts and Stability of Asymptotically Safe Gravity in the Minkowski Background}
\author{Changjun Gao$^{1,2}$}\email{gaocj@bao.ac.cn}
\author{Anzhong Wang$^{3}$}\email{anzhong_wang@baylor.edu}
\affiliation{$^{1}$The
National Astronomical Observatories, Chinese Academy of Sciences, Beijing 100012, China}
\affiliation{{$^{2}$Kavli Institute for Theoretical
Physics China, CAS, Beijing 100190, China }}
\affiliation{$^{3}$GCAP-CASPER, Physics Department, Baylor University, Waco, TX 76798-7316, USA}

\date{\today}

%%%%%%%%%%%%%%%%%%%%%%%%%%%%%%%%%%%%%%%%%%%%%%%%%%%%%%%%%%%%%%%%%%%%%%%%%%
\begin{abstract}
We investigate the problems of ghosts and stability  in the framework of asymptotically safe theory of gravity in the Minkowski
background. Within one-loop corrections,  we obtain explicitly the constraints on the coupling constants. Applying them to
the ones recently-obtained  at the fixed point, we find that the corresponding theory is both ghost-free and stable. Our results
can be easily generalized to high-loop corrections.
\end{abstract}

% insert suggested PACS numbers in braces on next line
% insert suggested keywords - APS authors don't need to do this
%\keywords{}

\maketitle

%%%%%%%%%%%%%%%%%%%%%%%%%%%%%%%%%%%%%%%%%%%%%%%%%%%%%%%%%%%%%%%%%%%%%%%%%%
\section{Introduction}
In 1974, $^{^{\textrm{\textbf{,}}}}$t Hooft and Veltman \cite{hooft:74} showed that General Relativity is
notoriously non-renormalizable at the quantum level. Further extensive investigations from other researchers
\cite{deser:2,deser:3,deser:4,deser:5,deser:6,deser:7} confirmed this fact. In view of it,
Weinberg proposed an effective quantum theory of gravity that turns out to be ultraviolet complete and nonperturbatively
renormalizable, because it is endowed with asymptotically safe property \cite{wein:77}. It is found that the
renormalization group flows have a stable fixed point in the ultraviolet limit and a finite dimensional
critical surface of trajectories appears in the vicinity of this fixed point. This is very interesting and the theory has attracted
extensive studies \cite{wein:2,wein:3,wein:4,wein:5,wein:6,wein:7,wein:8,wein:9,cai:2010}. Very recent investigations \cite{wein:10,wein:11,wein:12,wein:13,wein:14,wein:15} reveal that the ultraviolet critical
surface may be only three dimensional even for truncations of the exact renormalization group equations which may have more
than three coupling constants. There are two classes of stable fixed points. If the coupling constants vanish at the fixed point, then the fixed point is called Gaussian. On the other hand, if the coupling constants do not vanish at the fixed point, then the point is called non-Gaussian.
At the non-Gaussian fixed point, all the coupling constants are  non-zero. So the non-Gaussian fixed point
is very much physically important. It is found that with the approaching of trajectories to the fixed point,
the energy-dependence property of the coupling constants becomes more and more negligible. Finally, the trajectories of motion are irrelevant to the cutoff energy scale.

The aim of this paper is to investigate the ghosts and stability for the asymptotically safe gravity theory
with the specific non-Gaussian point. We derive the constraints at the non-Gaussian point that the theory is required to be ghost-free and stable.
The paper is organized as follows. Section II presents a brief review of asymptotical safety.
In section III, we investigate the problems of ghosts and stability for the asymptotically safe gravity in the Minkowski background. in Section VI we present
our main  conclusions. We shall use the system of units with $G=c=\hbar=k_{B}=1$
and the metric signature $(-,\ +,\ +,\ +)$ throughout the paper.

\section{Brief review of Asymptotically safe gravity}

Taking account of multi-loop quantum corrections, the effective action for asymptotically safe theory of gravity  takes the form
\begin{eqnarray}
\label{eq:action}
S&=&\int d^4x\sqrt{-g}\left[\Lambda^4g_0(\Lambda)+\Lambda^2g_1(\Lambda)R+g_2(\Lambda)R_{\mu\nu}R^{\mu\nu}\right.\nonumber
\\
&&\left.+g_3(\Lambda)R^2
+g_4(\Lambda)R_{\mu\nu\rho\sigma}R^{\mu\nu\rho\sigma}
+\Lambda^{-2}g_5R\square R %+\Lambda^{-2}g_5R^3
\right.\nonumber
\\
&&\left.+\Lambda^{-2}g_6\left(\nabla_{\mu}R_{\alpha}^{\beta}\right)\left(\nabla^{\mu}R^{\alpha}_{\beta}\right)+\cdot\cdot\cdot\right]\;.
~~~~
\end{eqnarray}
Here $g$ denotes the determinant of the metric tensor $g_{\mu\nu}$, $R$ the Ricci scalar, $R_{\mu\nu}$ the Ricci
tensor, $R_{\mu\nu\rho\sigma}$ the Riemann tensor, $\square\equiv \nabla_{\mu}\nabla^{\mu}$ the four dimensional $\textrm{d}{'}$Alembertian operator and $\Lambda$ the momentum cutoff.  The coefficients $g_i$ (i = 0, 1, 2, 3) are all dimensionless
coupling constants which are assumed to be momentum dependant. In particular,
for small momentum we have
\begin{eqnarray}
\label{eq:our}
\Lambda^4g_0(\Lambda)=\frac{\lambda(\Lambda)}{8\pi}\;, \ \  \Lambda^2g_1(\Lambda)=\frac{1}{16\pi G(\Lambda)}\;, \ \
\end{eqnarray}
where $\lambda(\Lambda)$ and $G(\Lambda)$ correspond to the Einstein cosmological and Newtonian constants, respectively.
These coupling constants satisfy the following renormalization group equations:
\begin{eqnarray}
\Lambda \frac{d}{d\Lambda}g_n(\Lambda)=\beta_n[g_i(\Lambda)]\;.
\end{eqnarray}

The condition for a fixed point at $g_n=g_{n*}$ is that $\beta_n(g_{*})=0$ for arbitrary $n$.
On the other hand, the conditions for the coupling constants to be attracted to a fixed point at
$\Lambda=\infty$ can be achieved by investigating the behavior of $g_{n}(\Lambda)$ when it is very close to $g_{n*}$.
Assuming that $\beta_{n}$ is an analytic function in the vicinity of $g_{n*}$, we can then expand them in Taylor series:
\begin{eqnarray}
\beta_n(g)=\sum_m \left(\frac{\partial\beta_{n}}{\partial g_{m}}\right)_{*}\left(g_m-g_{*m}\right)\;.
\end{eqnarray}
So, the solution of renormalization group equation is given by
\begin{eqnarray}
g_n(\Lambda)=g_{*n}+\sum_N u_{n}^{N}\Lambda^{\zeta_N}\;,
\end{eqnarray}
where $u^N$ and $\zeta_{N}$ are the eigenvectors and eigenvalues of the matrix $B_{nm}$:
\begin{eqnarray}
\sum_m B_{nm}u_{m}^{N}=\zeta_{N}u_{n}^{N}\;.
\end{eqnarray}

Physically, it is required that the real part of the eigenvalues be negative such that the coupling constants
can approach the fixed point. In Ref.~\cite{wein:12}, by using the exact renormalization group equations, a non-Gaussian fixed point
is found  to be
 \begin{eqnarray}
 \label{eq:result}
g_{0*}&=&0.0042\;, \ \ g_{1*}=-0.0101\;,\nonumber\\ \ \ g_{2*}&=&-0.010\;,\ \  g_{3*}=0.0109\;.
\end{eqnarray}
The goal of this paper is to show, with this new results, there is no ghosts and tachyon for the asymptotically safe gravity in the background of the Minkowski spacetime.
We derive the constraints on the coupling constants that the theory is ghost-free and stable.

\section{Ghosts and Stability }

For our purpose, let's first   consider  one-loop corrections, and later we shall generalize our results to high-loop cases. The effective action for asymptotically safe gravity
with one-loop corrections is given by,
\begin{eqnarray}
\label{eq:action}
S&=&\int d^4x\sqrt{-g}\left(\Lambda^2g_1R+g_2R_{\mu\nu}R^{\mu\nu}+g_3R^2\right)\;.
\end{eqnarray}
Since we are going to work in the Minkowski  background, we are forced to put $g_0=0$. On the other hand,
due to the Gauss-Bonnet theorem, we can express $R_{\mu\nu\rho\sigma}R^{\mu\nu\rho\sigma}$ as the linear combination of $R^2$ and $R_{\mu\nu}R^{\mu\nu}$. So for the one-loop action, it is sufficient to consider only the $g_2$ and $g_3$ terms.
In the Minkowski background, the metric takes the form,
\begin{eqnarray}
ds^2&=&-\left(1+2\varepsilon\Phi\right)dt^2+\left(1-2\varepsilon\Psi\right)dx^{i}dx_{i}\;,
\end{eqnarray}
where $\Phi$ and $\Psi$ are the gravitational potentials. It is convenient to introduce the dimensionless constant $\varepsilon$ which actually represents the order of perturbations.
In order to find the condition to be   ghost-free, we should
work out the perturbations to the second order in terms of $\Phi$ and $\Psi$.
Then,  we have the following expressions:
\begin{eqnarray}
\label{eq:quantities}
\sqrt{-g}&=&1+\varepsilon\left(\Phi-3\Psi\right)+\varepsilon^2\left(\frac{3}{2}\Psi^2-3\Phi\Psi-\frac{1}{2}\Phi^2\right)\;,\nonumber\\
R&=&-\varepsilon\left(2\nabla^2\Phi+6\ddot{\Psi}-4\nabla^2\Psi\right)
+\varepsilon^2\left[4\nabla^2\Phi\left(\Phi-\Psi\right)\right.\nonumber
\\
&&\left.+16\Psi\nabla^2\Psi+2\left(\nabla\Phi\cdot\nabla\Phi+\nabla\Phi\cdot\nabla\Psi\right)
\right.\nonumber
\\
&&\left.+12\ddot{\Psi}\left(\Phi-\Psi\right)+6\left(\dot{\Phi}\dot{\Psi}+\nabla\Psi\cdot\nabla\Psi\right)\right]\;,\nonumber\\
R^2&=&\varepsilon^2\left(2\nabla^2\Phi+6\ddot{\Psi}-4\nabla^2\Psi\right)^2\;,\nonumber\\
G^0_0&=&-2\varepsilon\nabla^2\Psi\;,\nonumber\\
G^0_i&=&-2\varepsilon\nabla_i\dot{\Psi}\;,\nonumber\\
G^i_j&=&\varepsilon\left[\left(\nabla^i\nabla_j-\frac{1}{3}\delta^i_j\nabla^2\right)\left(\Psi-\Phi\right)\right.\nonumber
\\
&&\left.+\frac{2}{3}
\delta^i_j\left(3\ddot{\Psi}+\nabla^2\Phi\right)\right]\;,\nonumber\\
R^{\mu}_{\nu}R^{\nu}_{\mu}&=&G^{\mu}_{\nu}G^{\nu}_{\mu}\;\nonumber\\
&=&\varepsilon^2\left[4\left(\nabla^2\Psi\right)^2-8\nabla\dot{\Psi}\cdot\nabla\dot{\Psi}\right]\nonumber\\&&
+\varepsilon^2\left\{\nabla_i\nabla_j\left(\Psi-\Phi\right)\nabla^i\nabla^j\left(\Psi-\Phi\right)
\right.\nonumber
\\
&&\left.-\frac{2}{3}\nabla^2\left(\Psi-\Phi\right)
\left[\nabla^2\left(\Psi-\Phi\right)-2\left(3\ddot{\Psi}+\nabla^2\Phi\right)\right]
\right.\nonumber\\&&\left.+\frac{1}{3}\left[\nabla^2\left(\Psi-\Phi\right)-
2\left(3\ddot{\Psi}+\nabla^2\Phi\right)\right]^2\right\}\;.
\end{eqnarray}
Here $\nabla^2\equiv\delta^{ij}\partial_{i}\partial_{j}$ represents the three dimensional Laplace operator and $i,\ j=1,\ 2,\ 3$.
Substituting the above expressions into the action, Eq.~(\ref{eq:action}), and integrating by parts, we obtain
the Lagrangian density as follows:
\begin{eqnarray}
\label{eq:L}
\mathscr{L}&=&-\Lambda^2g_1\varepsilon^2\left(4\nabla\Phi\cdot\nabla\Psi+6\dot{\Psi}^2-2\nabla\Psi\cdot\nabla\Psi\right)\nonumber\\
&&+g_2\varepsilon^2\left\{4\left(\nabla^2\Psi\right)^2-8\nabla\dot{\Psi}\cdot\nabla\dot{\Psi}+\frac{4}{3}\left(3\ddot{\Psi}+\nabla^2\Phi
\right)^2
\right.\nonumber
\\
&&\left.+\frac{2}{3}\left[\nabla^2\left(\Psi-\Phi\right)\right]^2\right\}\nonumber
\\&&+g_3\varepsilon^2\left(2\nabla^2\Phi+6\ddot{\Psi}-4\nabla^2\Psi\right)^2\;.
\end{eqnarray}

\subsection{Ghost-free Constraints}

It is convenient to consider the low energy and high energy limits of the Lagrangian density separately, which are given, respectively, by
\begin{eqnarray}
\label{eq:Low}
\mathscr{L}_L&=&-\Lambda^2g_1\varepsilon^2\left(4\nabla\Phi\cdot\nabla\Psi+6\dot{\Psi}^2-2\nabla\Psi\cdot\nabla\Psi\right), ~~~~
\end{eqnarray}
 and
\begin{eqnarray}
\label{eq:High}
\mathscr{L}_H&=&g_2\varepsilon^2\left\{4\left(\nabla^2\Psi\right)^2-8\nabla\dot{\Psi}\cdot\nabla\dot{\Psi}+\frac{4}{3}\left(3\ddot{\Psi}+\nabla^2\Phi
\right)^2
\right.\nonumber
\\
&&\left.+\frac{2}{3}\left[\nabla^2\left(\Psi-\Phi\right)\right]^2\right\}\nonumber
\\&&+g_3\varepsilon^2\left(2\nabla^2\Phi+6\ddot{\Psi}-4\nabla^2\Psi\right)^2\;.
\end{eqnarray}
To have the theory be free of  ghosts in all the energy scales,   the kinetic terms in both $\mathscr{L}_L$ and $\mathscr{L}_H$  must be positive. This
yields,
\begin{eqnarray}
g_1<0\;, \ \ \ g_2+3g_3>0\;.
\end{eqnarray}
%These are the requirements for ghost-free of the theory.

\subsection{Constraints from Stability}

In order to investigate the stability problem, we shall first derive the equation of motion for the scalar potential $\Phi$
and $\Psi$. To this end, let's substitute   Eq.~(\ref{eq:L}) into the Euler-Lagrange equation, then we  find
\begin{eqnarray}
\label{eq:Phi}
&&-4\Lambda^2g_1\nabla^2\Psi+g_2\left(\frac{4}{3}\nabla^4\Psi-8\nabla^2\ddot{\Psi}-4\nabla^4\Phi\right)\nonumber\\&&
+g_3\left(16\nabla^4\Psi-8\nabla^4\Phi-24\nabla^2\ddot{\Psi}\right)=0\;,
\end{eqnarray}
and
\begin{eqnarray}
\label{eq:Psi}
&&g_2\left(\frac{28}{3}\nabla^4\Psi-16\nabla^2\ddot{\Psi}+24\ddddot
{\Psi}+8\nabla^2\ddot{\Phi}-\frac{4}{3}\nabla^4\Phi\right)\nonumber\\&&
+g_3\left(72\ddddot{\Psi}+32\nabla^4\Psi-16\nabla^4\Phi+24\nabla^2\ddot{\Phi}-96\nabla^2\ddot{\Psi}\right)\nonumber\\&&
-4g_1\left(\nabla^2\Psi-3\ddot{\Psi}-\nabla^2\Phi\right)=0\;.
\end{eqnarray}

From Eq.~(\ref{eq:Phi}) we obtain
\begin{eqnarray}
\label{eq:phi}
\nabla^4\Phi&=&\frac{1}{g_2+2g_3}\left[-g_1\nabla^2\Psi+g_2\left(\frac{1}{3}\nabla^4\Psi-2\nabla^2\ddot{\Psi}\right)
\right.\nonumber
\\
&&\left.+g_3\left(4\nabla^4\Psi-6\nabla^2\ddot{\Psi}\right)
\right]\;.
\end{eqnarray}
Substituting this equation into $\nabla^2\otimes \textrm{Eq}.~(\ref{eq:Psi})$, we obtain the equation of motion for $\Psi$.
Then, considering the $k$ mode of its Fourier transformation:
 \begin{eqnarray}
\Psi=\int d^3k d\omega\Psi_k\left(\omega,\ \vec{k}\right)e^{i\left(\omega t-\vec{k}\cdot\vec{x}\right)}\;,
\end{eqnarray}
we obtain
\begin{eqnarray}
&&-\lambda^2g_1\left\{4k^4-12\omega^2k^2+\frac{4}{g_2+2g_3}\left[-\lambda^2g_1k^2\right.\right.\nonumber\\
& & \left.\left. -g_2\left(\frac{k^4}{3}-2\omega^2k^2\right)\nonumber
%\\&&\left.\left.
+g_3\left(6\omega^2k^2-4k^4\right)\right]\right\}\nonumber\\&&
+g_2\left(8k^4\omega^2-24k^2\omega^4-8k^6\right)\nonumber\\&&
+g_3\left(72k^4\omega^2-72\omega^4k^2-16k^6\right)=0\;.
\end{eqnarray}
In the low energy limit, terms  proportional to $g_2$ and $g_3$ are negligible, and the above equation reduces to
\begin{eqnarray}
g_1k^2=0\;,
\end{eqnarray}
which is exactly the result of Poisson equation for Newtonian gravity in Minkowski spacetime:
\begin{eqnarray}
\nabla^2\Psi=0\;.
\end{eqnarray}
In the high energy limit, terms proportional to $g_1$ are negligible, and the above equation turns out to be
\begin{eqnarray}
\label{eq:second}
\omega^4+Bk^2\omega^2+Ak^4=0,
\end{eqnarray}
where $B$ and $A$ are defined as
\begin{eqnarray}
B&=&-\frac{g_2+9g_3}{3\left(g_2+3g_3\right)}\;,\nonumber\\
A&=&\frac{g_2+2g_3}{3\left(g_2+3g_3\right)}\;.
\end{eqnarray}
Then we obtain from Eq.~(\ref{eq:second})
\begin{eqnarray}
\omega_{\pm}^2=\frac{k^2}{2}\left(-B\pm\sqrt{B^2-4A}\right)\;.
\end{eqnarray}
%The stability requires  $\omega_{\pm}^{2} >0$.
To study the  stability problem, it is found convenient to consider the cases, $B = 0,\; B > 0$, and $B< 0$ separately.

\subsubsection{$B=0$}

When $B = 0$, we  divide it further into   five sub-cases.

1) $B=0,\ A>0$: This implies that
\begin{eqnarray}
\label{eqa}
g_3>0\;,\ \ \ g_2=-9g_3\;,
\end{eqnarray}
or
\begin{eqnarray}
g_3<0\;,\ \ \ g_2=-9g_3\;.
\end{eqnarray}
For the first solution given by Eq.(\ref{eqa}), it can be shown that there is no ghost. But the corresponding
theory is unstable,  since now we have $\bigtriangleup\equiv B^2-4A<0$.
For the second solution, there are both ghost and instability in the theory.

2) $B=0,\ A=0$: This is   a trivial case, in which we have
\begin{eqnarray}
g_2=0\;,\ \ \ g_3=0\;.
\end{eqnarray}
Thus, the corresponding theory is Einstein's.

3) $B=0,\ A<0$: In this case, it can be shown that there is no solution.

\subsubsection{$B>0$}

Similar to the last case, we further divide it into   the following sub-cases.

1) $B>0,\ A>0,\ \triangle>0$: Then, we find the solutions are given by,
\begin{eqnarray}
g_3>0\;,\ \ \ -\frac{3}{11}\left(7+2\sqrt{15}\right)g_3<g_2<-3g_3\;,
\end{eqnarray}
or
\begin{eqnarray}
g_3<0\;,\ \ \ -\frac{3}{11}\left(7+2\sqrt{15}\right)g_3>g_2>-3g_3\;.
\end{eqnarray}
For the first solution, there is no ghost. But the theory is unstable, since $\omega$ is pure imaginary.
For the second solution, there are both ghost and instability.

2) $B>0,\ A>0,\ \triangle=0$: In this case, we find the solution is given by,
\begin{eqnarray}
g_3>0\;,\ \ \ g_2=-\frac{3}{11}\left(7+2\sqrt{15}\right)g_3\;.
\end{eqnarray}
Then, there is no ghost. But the theory is unstable since $\omega$ is pure imaginary.

3) $B>0,\ A>0,\ \triangle<0$:  We find the solution is given by:
\begin{eqnarray}
g_3>0\;,\ \ \ -9g_3<g_2<-\frac{3}{11}\left(7+2\sqrt{15}\right)g_3\;.
\end{eqnarray}
In this case, there is no ghost. But the theory is unstable because of $\triangle<0$.

4) $B>0,\ A=0$: It can be shown that in this case there is no solution.

5) $B>0,\ A<0$: We find the solutions are given by:
\begin{eqnarray}
g_3>0\;,\ \ \ -3g_3<g_2<-2g_3\;,
\end{eqnarray}
or
\begin{eqnarray}
g_3<0\;,\ \ \ -3g_3>g_2>-2g_3\;,
\end{eqnarray}
For the first solution, there is no ghost. Furthermore, the theory is stable.
For the second solution, the theory is stable but not ghost-free.

\subsubsection{$B<0$}

In this case, we can also divide it into five sub-cases.

1) $B<0,\ A>0,\ \triangle>0$: Then,  we find the solutions are given by,
\begin{eqnarray}
g_3>0\;,\ \ \ -\frac{3}{11}\left(7-2\sqrt{15}\right)g_3>g_2>-2g_3\;,
\end{eqnarray}
or
\begin{eqnarray}
g_3<0\;,\ \ \ -\frac{3}{11}\left(7-2\sqrt{15}\right)g_3<g_2<-2g_3\;.
\end{eqnarray}
For the first solution, the theory is both stable and ghost-free.
For the second solution, the theory is stable but is not ghost-free.

2) $B<0,\ A>0,\ \triangle=0$: In this case, we find the solutions are given by,
\begin{eqnarray}
g_3>0\;,\ \ \ g_2=-\frac{3}{11}\left(7-2\sqrt{15}\right)g_3\;,
\end{eqnarray}
or
\begin{eqnarray}
g_3<0\;,\ \ \ g_2=-\frac{3}{11}\left(7+2\sqrt{15}\right)g_3\;.
\end{eqnarray}
For the first solution, the theory is both stable and ghost-free.
For the second solution, the theory is stable but is not ghost-free.

3) $B<0,\ A>0,\ \triangle<0$: Then, we find the solutions are given by:
\begin{eqnarray}
g_3>0\;,\ \ \ g_2<-9g_3\;,
\end{eqnarray}
\begin{eqnarray}
g_3>0\;,\ \ \ g_2>-\frac{3}{11}\left(7-2\sqrt{15}\right)g_3\;,
\end{eqnarray}
\begin{eqnarray}
g_3<0\;,\ \ \ g_2>-9g_3\;,
\end{eqnarray}
and
\begin{eqnarray}
g_3<0\;,\ \ \ g_2<-\frac{3}{11}\left(7-2\sqrt{15}\right)g_3\;.
\end{eqnarray}
Among the four solutions, two of them are ghost-free. But the four solutions are all unstable because of
$\triangle<0$.

4) $B<0,\ A=0$: In this case, we find the solutions are given by,
\begin{eqnarray}
g_3>0\;,\ \ \ g_2=-2g_3\;,
\end{eqnarray}
or
\begin{eqnarray}
g_3>0\;,\ \ \ g_2=-2g_3\;.
\end{eqnarray}
For the first solution, the theory is both stable and ghost-free.
For the second solution, the theory is stable but is not ghost-free.

5) $B<0,\ A<0$: Then,  we find the solutions are given by,
\begin{eqnarray}
g_3>0\;,\ \ \ -3g_3<g_2<-2g_3\;,
\end{eqnarray}
and
\begin{eqnarray}
g_3<0\;,\ \ \ -2g_3<g_2<-3g_3\;.
\end{eqnarray}
For the first solution, the theory is both stable and ghost-free.
For the second solution, the theory is stable but is not ghost-free.

Combining all the above discussions, we obtain the following constraints for the theory to be both
ghost-free and stable,
\begin{eqnarray}
\label{cnds}
&&g_1<0\;,\nonumber\\&& g_3>0\;,\nonumber\\ & & -3g_3<g_2\leq-\frac{3}{11}\left(7-2\sqrt{15}\right)g_3\;.
\end{eqnarray}

Turning  to the numerical results, Eq.~(\ref{eq:result}), which are related our constants as %ollows:
$g_{i} = g_{i *},\; (i = 1, 2, 3)$,
%&g_1=g_{1*}\;,\nonumber\\&& g_2=g_{2*}\;,\nonumber\\ &&g_3=g_{3*}\;,
%end{eqnarray}
we find that  Eq.~(\ref{eq:result}) does satisfy
the above constraints.

\section{conclusion and discussion}

In this paper, we have investigated the problems of ghosts and
stability in the framework of the asymptotically safe theory of
gravity, and found explicitly the conditions (\ref{cnds}), with
which the theory is ghost-free and stable. These conditions are
satisfied by  the newly-obtained values  of the coupling constants
\cite{wein:12}, given by Eq.(\ref{eq:result}) at the fixed point.
Therefore,  it is concluded that the asymptotical safe theory of
gravity is both  stable and ghost-free. Actually,
Ref.~\cite{dvali:2010} argues that the asymptotic gravity contains
ghosts only if the gravitational constant decays in the weak
regime. On the contrary, if the gravitational constant does not
decay in the weak regime there are probably no ghost.

It should be noted that strictly speaking our above conclusion is
true only for the effective action with  one loop corrections. In
the following, we shall argue that the conditions
 (\ref{cnds}) equally hold when high loop corrections are taken into account. To show this clearly, let us first concentrate ourselves on the second loop corrections, which add
 nine independent six-order derivative terms to the effective action \cite{deser:3}, and in Eq.(\ref{eq:action}) we only presented explicitly two of them, represented by the
 dimensionless coupling constants  $g_{5}$ and $g_{6}$, because these are the only terms that are relevant to the
ghost and instability problems. All the  rest is of third order of $\varepsilon$, as one can see from the expressions given by Eq.(\ref{eq:quantities}), in which we find
that $R_{ij} \simeq {\cal{O}}(\varepsilon)$ in the Minkowski background. Therefore, they have no contributions to  the quadratic action. The $g_{5, \;6}$ terms, on the other
hand, are six-order derivatives, which are dominant only when the energy is sufficiently high, say, higher than $E_{6th}$, that is, $\omega,\;  k \gg E_{6th}$. In such a high
 energy scale,  the requirements for the theory to be ghost-free and stable will impose conditions only on the coupling constants $g_{5}$ and $g_{6}$, as the relevant terms
 proportional to $g_{1,\; 2, \; 3}$ are all negligible, provided that no fine-tuning exists. As the energy decreases,  these terms become less and less important, and finally the
 terms proportional to $g_{2}$ and $g_{3}$
 become dominant. Then, the conditions for the theory to be free of ghosts and instability are given precisely by Eq.(\ref{cnds}) for $g_{2}$ and $g_{3}$. As the energy
 continuously decreases,  the $g_{1}$ terms will finally become dominant, and these requirements then lead to the condition $g_1 < 0$, as given by Eq.(\ref{cnds}).
 For the nth loop, only the quadratic terms like
 $R\Box^{(n-1)}R$  are relevant    to the ghost and instability problems, which are the  polynomials of the $2(n+1)$-order of $\omega$ and $k$.
  Again, these terms become dominant only at even higher energy scale. Following the same arguments as given above, one can see that they have no effects on the low
  energy constants, as far as the  ghost- and
 instability-free conditions are concerned. As a result, the conditions (\ref{cnds}) hold even when high order corrections are taken into account.

\acknowledgments We thank Dr. Cristiano Germani and Dr. Yi-Fu Cai
for helpful communications. The author C.G. gratefully
acknowledges the support of K. C. Wong Education Foundation, Hong
Kong and the NSFC under Grant No.10525314, 10533010, 10575004,
10973014, by the Chinese Academy of Sciences KJCX3-SYW-N2, and by
the 973 Projects No. 2007CB815401 and 2010CB833004. The work of
A.W. was supported in part by DOE Grant, DE-FG02-10ER41692. C.G.
acknowledges the hospitality of the Baylor Physics Department
where this work was carried out.

\newcommand\ARNPS[3]{~Ann. Rev. Nucl. Part. Sci.{\bf ~#1}, #2~ (#3)}
\newcommand\AL[3]{~Astron. Lett.{\bf ~#1}, #2~ (#3)}
\newcommand\APP[3]{~Astropart. Phys.{\bf ~#1}, #2~ (#3)}
\newcommand\AP[3]{~Ann. Phys.{\bf ~#1}, #2~ (#3)}
\newcommand\AJ[3]{~Astron. J.{\bf ~#1}, #2~(#3)}
\newcommand\APJ[3]{~Astrophys. J.{\bf ~#1}, #2~ (#3)}
\newcommand\APJL[3]{~Astrophys. J. Lett. {\bf ~#1}, L#2~(#3)}
\newcommand\APJS[3]{~Astrophys. J. Suppl. Ser.{\bf ~#1}, #2~(#3)}
\newcommand\JHEP[3]{~JHEP{\bf ~#1}, #2~(#3)}
\newcommand\JCAP[3]{~JCAP {\bf ~#1}, #2~ (#3)}
\newcommand\LRR[3]{~Living Rev. Relativity. {\bf ~#1}, #2~ (#3)}
\newcommand\MNRAS[3]{~Mon. Not. R. Astron. Soc.{\bf ~#1}, #2~(#3)}
\newcommand\MNRASL[3]{~Mon. Not. R. Astron. Soc.{\bf ~#1}, L#2~(#3)}
\newcommand\NPB[3]{~Nucl. Phys. B{\bf ~#1}, #2~(#3)}
\newcommand\PLB[3]{~Phys. Lett. B{\bf ~#1}, #2~(#3)}
\newcommand\PRL[3]{~Phys. Rev. Lett.{\bf ~#1}, #2~(#3)}
\newcommand\PR[3]{~Phys. Rep.{\bf ~#1}, #2~(#3)}
\newcommand\PTP[3]{~Prog. Theor. Phys.{\bf ~#1}, #2~(#3)}
\newcommand\PRD[3]{~Phys. Rev. D{\bf ~#1}, #2~(#3)}
\newcommand\RMP[3]{~Rev. Mod. Phys.{\bf ~#1}, #2~(#3)}
\newcommand\SJNP[3]{~Sov. J. Nucl. Phys.{\bf ~#1}, #2~(#3)}
\newcommand\ZPC[3]{~Z. Phys. C{\bf ~#1}, #2~(#3)}
 \newcommand\IJGMP[3]{~Int. J. Geom. Meth. Mod. Phys.{\bf ~#1}, #2~(#3)}
 \newcommand\IJMPA[3]{~Int. J. Mod. Phys. A{\bf ~#1}, #2~(#3)}
  \newcommand\MPLA[3]{~Mod. Phys. Lett. A{\bf ~#1}, #2~(#3)}
  \newcommand\GRG[3]{~Gen. Rel. Grav.{\bf ~#1}, #2~(#3)}
   \newcommand\CQG[3]{~Class. Quan. Grav.{\bf ~#1}, #2~(#3)}
   \newcommand\APPTA[3]{~Annales. Poincare. Phys. Theor. A{\bf ~#1}, #2~(#3)}
    \newcommand\RPJ[3]{~Russ. Phys. J.{\bf ~#1}, #2~(#3)}
   \newcommand\IVF[3]{~ Izv. VUZ, Fiz. {\bf ~#1}, #2~(#3)}

\end{document}